\documentclass{aa}
\usepackage{epsfig}

\def\power#1{\mbox{$\times10^{#1}\ $}}
\newcommand{\gap}{\mathrel{ \rlap{\raise.5ex\hbox{$>$}}
                      {\lower.5ex\hbox{$\sim$}}  } }
\newcommand{\lap}{\mathrel{ \rlap{\raise.5ex\hbox{$<$}}
		      {\lower.5ex\hbox{$\sim$}}  } }

\begin{document}

\thesaurus{99     %
              (99.999)} 
\title{Big Bang Nucleosynthesis updated with the NACRE Compilation}

\author{Elisabeth Vangioni-Flam\inst{1}, Alain Coc\inst{2} and 
Michel Cass\'e\inst{1, 3}}
          
\institute{
Institut d'Astrophysique de Paris, 98 bis Bd Arago 75014 Paris, France e-mail:flam@iap.fr
\and
Centre de Spectrom\'etrie Nucl\'eaire et de Spectrom\'etrie
de Masse, IN2P3-CNRS and Universit\'e Paris Sud, B\^atiment 104,\\ 
91405 Orsay Campus, France
\and
Service d'Astrophysique, DAPNIA, DSM, CEA, Orme des Merisiers,
91191 Gif sur Yvette CEDEX France
}

\date{Received ....; Accepted ....}

\maketitle

\begin{abstract}
We update the Big Bang Nucleosynthesis calculations on the basis of the recent
NACRE compilation of reaction rates. The average values of the calculated
abundances of light nuclei do not differ significantly from those obtained
using the previous Fowler's compilation.
However,  ${^7}Li$ is slightly depressed at high baryon to photon ratio $\eta$.
Concerning ${^{10}}B$, its abundance is significantly lower than the one
calculated with the Caughlan and Fowler (1988) rates as anticipated by
Rauscher and Raimann (1997).
We estimate the uncertainties related to the nuclear reaction
rates on the abundances of $D$, ${^3}He$, ${^4}He$, ${^6}Li$, ${^7}Li$,
${^9}Be$, ${^{10}}B$ and ${^{11}}B$ of cosmological and astrophysical interest.
The main uncertainty concerns the $D(p,\gamma){^3}He$  reaction rate affecting
the synthesis of ${^7}Li$ 
at rather high baryonic density and also 
 the ${^3}He(\alpha,\gamma){^7}Be$ and ${^7}Li(p,\alpha){^4}He$
 reactions.
On the left part of the lithium valley the uncertainty is reduced
due to the improvement of the measurement of the $T(\alpha,\gamma)^{7}Li$
reaction rate.
The observed abundances of the nuclei of interest are compared to the
predictions of the BBN model, taking into account both observational
and theoretical uncertainties.
Indeed, the ${^7}Li$ abundance observed in halo stars (Spite
plateau) is now  determined with high precision since the thickness of
this plateau appears, in the light of recent observations,
exceptionnaly small ($<$ 0.05 dex). The potential destruction/dilution of
${^7}Li$ in the outer layers of halo stars
which could mask the true value of the primordial abundance is in full debate,
but the present trend is towards a drastic reduction of the depletion factor
(about 0.10 dex).
It is why we use this isotope as a preferred baryometer. Even though much
efforts have been devoted to the determination of deuterium in absorbing
clouds in the line of sight of remote quasars, the statistics is very poor
compared to the long series of lithium measurements.
Taking into account these lithium constraints, two possible baryonic density
ranges emerge, $\eta_{10}= 1.5 - 1.9$ and $\eta_{10} = 3.3 - 5.1$.
In the first case, $Li$ is in concordance with $D$ from Webb et al (1997) and
${^4}He$ from Fields and Olive (1998) and Peimbert and Peimbert (2000).
In the second case, agreement is achieved with $D$ from Tytler et al (2000)
and ${^4}He$ from Izotov and Thuan (1998). 

Concerning the less abundant light isotopes, the theoretical BBN abundance
of ${^6}Li$ is affected by a large uncertainty due to the poor knowledge
of the $D(\alpha,\gamma){^6}Li$ reaction rate. However, at high $\eta$,
its abundance is so low that there is little chance to determine
observationally the true BBN ${^6}Li$ abundance.
But, at low $\eta$, its abundance being one thousandth of that of primordial
${^7}Li$, 6/7 ratio measurements at very low metallicity are not totally
hopeless in the future.
Nevertheless, in the present situation, ${^6}Li$ is cosmologically relevant,
though indirectly, since its mere presence in a few halo stars, corroborates
the fact that it is essentially intact in these stars together with $^{7}Li$
and thus the Spite plateau can be used as such to infer the primordial
${^7}Li$ abundance.
The $Be$ and $B$ abundances produced in the Big Bang  are orders of magnitudes
lower, and spallation of fast carbon and oxygen is probably their unique
source, in the early Galaxy.   
\end{abstract}

\section{Introduction}
Besides the expansion of the Universe and the ubiquitous presence of the
fossil cosmological radiation, the Big Bang Nucleosynthesis (BBN) is one
pilar of modern cosmology.
It allows, in principle, the derivation of the baryonic density of the
universe (see for reviews Schramm and Turner 1998, Sarkar 1996, 1999 and
Olive et al 2000).
The determination of light element abundances has improved dramatically in
the recent past and the planned observations of $D$, ${^4}He$, ${^6}Li$
and ${^7}Li$ should allow a precise determination (10\% is a
reachable goal) of the universal baryonic density, provided the
precision of the model is made compatible with this objective.
Taking advantage of the  release of a new compilation of thermonuclear
reation rates, called NACRE (Nuclear Astrophysics Compilation
of REaction rates) (Angulo et al 1999), we have updated the standard
BBN model developed at the Institut d'Astrophysique de Paris, including
the analysis of the $Be$ and $B$ production in the Big Bang.

On the other hand, observations of light isotopes have florished:
i) $D/H$ has been observed in absorbing clouds on the sightline of remote 
quasars, ii) refined observations of $^{4}He$ have been performed in 
extragalactic very metal poor HII regions and in the Small Magellanic Cloud 
(Peimbert and Peimbert 2000), iii) the $^{6}Li$ abundance has been determined
in two halo stars, iv) high quality $^{7}Li$ observations in the halo stars
have been accumulated.
A review of the present data can be found in Olive et al (2000) and 
Tytler et al (2000).
Thus, it is timely to reassess the determination of the baryonic density of 
the Universe in the light of advances in nuclear physics and astronomical 
observations.

In section II, we present the new compilation of the reaction rates
and compare it with the classical Caughlan-Fowler (1988) ones; we evaluate
the sensitivity of the different light element abundances
to the change of each relevant reaction rate;
BBN calculations are performed using i) recommended values of the reaction 
rates and ii) extreme values obtained from the low and high rate limits.
In section III, we discuss the astrophysical status of each isotope, both
observationally and theoretically in order to confront it to the BBN
calculation; we deduce the baryonic density of the Universe.
Finally, we draw conclusion in section IV and stress the importance of
precise measurements of a few key nuclear cross sections and refined
abundance determinations of $D$ and $^{6}Li$.

\section{Nuclear Physics and Big Bang nucleosynthesis }

The new compilation of Angulo et al (1999) of thermonuclear reaction rates
for astrophysics, includes 86 charged-particle induced reactions
corresponding to the proton capture reactions involved in the cold
pp-chain, CNO cycle, NeNa and MgAl chains.
The BBN network is constituted by about 60
reactions which participate to primordial nucleosynthesis up
to $^{11}B$. 22 of these reactions are covered by the NACRE compilation.
The main innovative features of NACRE with respect to the former
compilation Caughlan and Fowler (1988) are the following: (1) detailed references
are provided to the source of original data; (2) uncertainties
are analyzed in detail, realistic lower and
upper bounds of the rates are provided; (3) the rates are given in tabular
form, available also electronically on the World--Wide--Web.
For these reasons, we can adopt the NACRE recommended rates for the calculation
of the yields and use the upper and lower limits of the rates to test
the sensitivity of the abundances to the nuclear uncertainties. As the origins
for these uncertainties are documented in Angulo et al (1999), we do not
discuss them here unless they show a significant effect on yields.
We calculated the isotopic abundances as a function of
$\eta_{10}$ between 1 and 10, changing one single reaction at a time.
For each reaction we made a calculation with the high and low NACRE limits
while the remaining reaction rates were set to their recommended NACRE value.
Then we calculated, the maximum of the quantity
${\Delta}N/N{\equiv}N_{high}/N_{low}-1$ within the range of $\eta_{10}$
variations for each of the 8 isotopes. Positive (resp. negative) values
correspond to higher (resp. lower) isotope production when the high rate is
used instead of the low one. Note however that following the ${\Delta}N/N$
definition, the positive values are not bound while the negative values are
bound by -1. Hence, for instance (see Table 1), ${\Delta}N/N=+10$ (resp.
-0.78) means that the isotope yield is 11 times higher (resp. 4.5 times lower)
with the high rate than with the low one. 
The corresponding results are shown in Table 1.

\begin{table}
\caption{Influencial reactions and their sensitivity to nuclear uncertainties}
\tiny
\begin{flushleft}
\begin{tabular}{|l|c|c|c|c|c|c|c|c|}
\hline
Reaction~${\backslash}$~${\Delta}N/N$
& $^4$He&D&$^3$He&$^7$Li&$^6$Li&$^9$Be&$^{10}$B&$^{11}$B\\
\hline
$^1$H(p,$e^+\nu)^2$H&n.s.&n.s.&n.s.&n.s.&n.s.&n.s.&
n.s.&n.s.\\
\hline
$^2$H(p,$\gamma)^3$He&n.s.&-0.19&0.19&0.26&-0.19&-0.27&-0.18&0.20\\
\hline
$^2$H(d,$\gamma)^4$He&n.s.&n.s.&n.s.&n.s.&n.s.&n.s.&
n.s.&n.s.\\
\hline
$^2$H(d,n)$^3$He&n.s.&-0.09&0.06&0.12&-0.09&-0.09&-0.08&0.19\\
\hline
$^2$H(d,p)$^3$H&n.s.&-0.03&-0.04&0.01&-0.03&0.04&-0.03&0.03\\
\hline
$^2$H($\alpha,\gamma)^6$Li&n.s.&n.s.&n.s.&n.s.&21.&n.s.
&10.&n.s.\\
\hline
$^3$H(d,n)$^4$He&n.s.&n.s.&n.s.&-0.07&n.s.&-0.13&-0.04&-0.07\\
\hline
$^3$H($\alpha,\gamma)^7$Li&n.s.&n.s.&n.s.&0.24&n.s.&0.24
&0.06&0.24\\
\hline
$^3$He($^3$He,2p)$^4$He&n.s.&n.s.&n.s.&n.s.&n.s.&n.s.
&n.s.&n.s.\\
\hline
$^3$He($\alpha,\gamma)^7$Be&n.s.&n.s.&n.s.&0.39&n.s.&0.21
&n.s.&0.40\\
\hline
2$^{4}$He($\alpha,\gamma)^{12}$C&n.s.&n.s.&n.s.&n.s.&n.s.&
n.s.&n.s.&0.01\\
\hline
2$^{4}$He(n,$\gamma)^{9}$Be&n.s.&n.s.&n.s.&n.s.&n.s.&n.s.
&n.s.&n.s.\\
\hline
$^6$Li(p,$\gamma)^7$Be&n.s.&n.s.&n.s.&n.s.&n.s.&n.s.
&n.s.&n.s.\\
\hline
$^6$Li(p,$\alpha)^3$He&n.s.&n.s.&n.s.&n.s.&
-0.18&n.s.&-0.18&n.s.\\
\hline
$^7$Li(p,$\gamma)^8$Be&n.s.&n.s.&n.s.&n.s.&n.s.&n.s.
&n.s.&n.s.\\
\hline
$^7$Li(p,$\alpha)^4$He&n.s.&n.s.&n.s.&-0.25&n.s.&-0.25&
-0.07&-0.24\\
\hline
$^7$Li($\alpha,\gamma)^{11}$B&n.s.&n.s.&n.s.&n.s.&n.s.&n.s.
&n.s.&0.39\\
\hline
$^{7}$Li($\alpha$,n)$^{10}$B&\multicolumn{8}{c|}{Q$<0$}\\
\hline
$^7$Be(p,$\gamma)^8$B&n.s.&n.s.&n.s.&n.s.&n.s.&n.s.&
n.s.&0.01\\
\hline
$^7$Be($\alpha,\gamma)^8$B&n.s.&n.s.&n.s.&n.s.&n.s.&n.s.&
n.s.&0.81\\
\hline
$^9$Be(p,$\gamma)^{10}$B&n.s.&n.s.&n.s.&n.s.&n.s.&n.s.&
0.04&n.s.\\
\hline
$^9$Be(p,pn)2$^4$He&n.s.&n.s.&n.s.&n.s.&n.s.&n.s.
&n.s.&n.s.\\
\hline
$^9$Be(p,d)2$^4$He&n.s.&n.s.&n.s.&n.s.&n.s.&-0.13
&-0.02&-0.01\\
\hline
$^9$Be(p,$\alpha)^6$Li&n.s.&n.s.&n.s.&n.s.&n.s.&-0.13
&-0.02&n.s.\\
\hline
$^9$Be($\alpha$,n)$^{12}$C&n.s.&n.s.&n.s.&n.s.&n.s.&n.s.
&n.s.&-0.01\\
\hline
$^{10}$B(p,$\gamma)^{11}$C&n.s.&n.s.&n.s.&n.s.&n.s.&n.s.
&n.s.&-0.01\\
\hline
$^{10}$B(p,$\alpha)^7$Be&n.s.&n.s.&n.s.&n.s.&n.s.&n.s.
&-0.38&n.s.\\
\hline
$^{11}$B(p,$\gamma)^{12}$C&n.s.&n.s.&n.s.&n.s.&n.s.&n.s.&
n.s.&n.s.\\
\hline
$^{11}$B(p,n)$^{11}$C&\multicolumn{8}{c|}{Q$<0$} \\
\hline
$^{11}$B(p,$\alpha)2^4$He&n.s.&n.s.&n.s.&n.s.&n.s.&n.s.
&n.s.&-0.78\\
\hline
$^{14}$N(p,$\alpha)^{11}$C&\multicolumn{8}{c|}{Q$<0$}\\
\hline
\end{tabular}
\newline
${\Delta}N/N{\equiv}N_h/N_l-1$
\newline
n.s. : not significant ($|{\Delta}N/N|<0.01$)
\newline
Q$<0$ : no tabulated reverse rates available for endoenergic reactions.
\end{flushleft}
\normalsize
\end{table}

\begin{figure} 
\epsfig{file=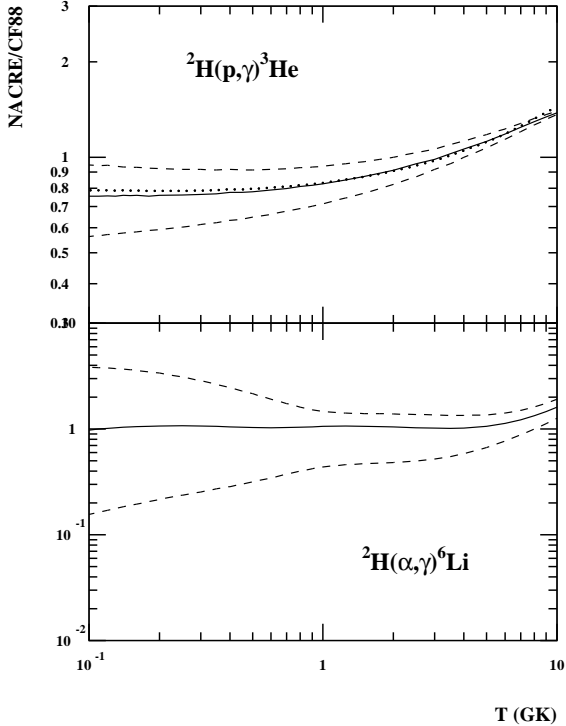,width=8.cm}
\vspace{10pt}
\caption{ Two particularly uncertain reaction rates: $D(p,\gamma){^3}He$
and $D(\alpha,\gamma){^6}Li$  (NACRE/Caughlan-Fowler 1988- CF88) solid line:
mean ratio and dashed line: NACRE upper limit/CF88 and NACRE lower limit/CF88.
The dotted line ($D(p,\gamma){^3}He$ reaction) represent the small effect of
the Schmidt et al. (1996) correction not included in the NACRE compilation.}
\end{figure}

For three of these reactions, the test has not been made because the NACRE
compilation does not provide high and low limits for the reverse rate of
endoenergic reactions ($^{7}Li(\alpha,n)^{10}B$ , 
$^{11}B(p,n)^{11}C$ and $^{14}B(p,\alpha)^{11}C$).
The reverse recommended rate can be calculated from
the formulas. However, the low and high rates are only tabulated and limited
down to temperatures chosen in order that the reaction rate remains above
the lower limit of
$N_{\rm A}\langle\sigma v\rangle \leq 10^{-25}$ cm$^3$ mol$^{-1}$ s$^{-1}$.
For $Q<0$ reactions, the reverse rate is higher than the direct tabulated
one but limited by $N_{\rm A}\langle\sigma v\rangle \leq 10^{-25}$ cm$^3$
mol$^{-1}$ s$^{-1}$ times the {\em reverse ratio}.

In the analysis of yield uncertainties, one should keep in mind that
the guidelines for the NACRE compilation favoured conservative
upper and lower limits for the rates in order that the actual rate be
within these limits with a high degree of confidence.
For instance, when incompatible data set were present, and if the differences
could not be resolved by analysing the publications alone the high and low
limits were set in order to incorporate all data sets.
In some case (e.g. $D(\alpha,\gamma)^6Li$) the incompatibility is between
experimental data and theoretical results making very problematic
 the interpretation of the
rate uncertainty in term of gaussian distributions. When only an upper
limit is available experimentally, as in previous compilations, its
contribution is weighted by a 0., 0.1 and 1. factor respectively.
This again makes difficult the probabilistic interpretation of rate 
uncertainties. 
Nevertheless, few NACRE reaction rates are at the origin of a significant 
uncertainty (Table 1).

The $D(p,\gamma)^3He$ NACRE reaction rate is responsible for a 20--30\%
uncertainty on most isotopes. It comes from the dispersion of experimental
results.
Note that the incompatibility of the two data set at low energy reported in
Angulo et al. (1999) has been removed after a correction factor 
(Schmidt et al. 1996) has been applied to Schmidt et al. (1995) data to 
account for an unsuspected experimental bias.
To check the effect of this rate update, we reiterate the NACRE calculation by 
fitting the experimental data points up to 2~MeV but with the corrected 
Schmidt et al. (1995,1996) data. Since it affects only the lowest energies,
it has a negligible effect in the domain of BBN (see Fig. 1). 

The most dramatic effect comes from the $D(\alpha,\gamma)^6Li$ reaction
which induces uncertainties of a factor 22 and 11 on the $^6Li$ and $^{10}$B
yields. This rate uncertainty originates from the discrepancy between
theoretical low energy dependance of the S--factor and experimental data
(Kiener et al. 1991) obtained with the coulomb break--up technique
(see Kharbach and Descouvemont (1998) for a recent comparison between 
theories and experiment).
This reaction clearly deserves further experimental effort.
The reactions  $^3He(\alpha,\gamma)^7Be$
and $^7Li(p,\alpha)^4He$ induce a significant (25--40\% each) uncertainty 
on $^7Li$ production. For these reactions, the rate uncertainties comes 
from the dispersion (systematic errors) of non resonant experimental 
data at low energy.

Uncertainties on $BeB$ isotope yields remain negligible when compared
with the gap between calculated values and observational limitations.
At maximum, a factor of $\approx4$ uncertainty on $^{11}B$ at
low $\eta$ arises from the influence of the $^{11}B(p,\alpha)^8Be$ reaction.
However, the NACRE compilation covers only 22 of the 60 reactions
involved in Big Bang Nucleosynthesis. In particular, $BeB$ yield 
uncertainties are most likely dominated by uncertainties in reaction 
rates not included in the NACRE compilation.

In front of the large systematic uncertainties on the observational abundance
data (section 3.1), it seems premature to elaborate complex 
statistical procedures to get very precise theoretical uncertainties on the
primordial abundances.

Indeed, extensive studies have used Monte-Carlo techniques to estimate the
theoretical uncertainties studies have used Monte-Carlo techniques 
(Krauss et al 1990, Smith et al 1993,  Fiorentini et al 1998, Olive et al 2000,
Nollet and Burles 2000). 
 These powerfull methods have proven their
 efficiency in various domains (e.g. simulations of high energy physics
 experiment) provided that the probability distribution involved are known.
 Concerning our approach, since the values given by the NACRE compilation
 do not represent statistical confidence level, but upper and lower limits,
 they are not directly appropriate to Monte-Carlo calculations, but they include
 both statistical and systematic errors (see above).

  We estimate the uncertainties performing to global calculations, one with
 all the reaction rates set to their lower limits, and the second
  one with all the reaction rates set to their higher limits
  (dashed lines in figures 4 and 5). This method could lead to compensation (according to
 the signs of individual uncertainties displayed in table 1) 
  between production and destruction and therefore to
 underestimate the global uncertainties in some cases. 
The advantage of this technique is simplicity and transparency. 
But the disadvantage is that it does not allow to derive a confidence level, 
in the statistical sense.

The primordial abundances of the light elements $D$, $^{3,4}He$ 
and $^{7}Li$ are governed by the expansion rate of the Universe and the 
cooling it induces.
Under the classical assumptions, (homogeneity and the isotropy of the Universe
and standard particle physics: three light neutrino species, neutron lifetime 
equal to 887 seconds) the abundances depend only on the baryon to photon ratio
$\eta$, related to the baryonic parameter by $\eta_{10} = 273\Omega_{B}.h^{2}$
with $h=H/100 kms^{-1}Mpc^{-1}$ (see e.g. for details Olive et al 2000).
 
We do not include the small corrections on the $^{4}He$ mass fraction due to
Coulomb, radiative and finite temperature effects, finite nucleon
mass effects and differential neutrino heating (Sarkar 1999, Lopez and 
Turner 1999) since these corrections lead to effects much less than the 
uncertainties on the observational data.

The network extends up to $^{11}C$ (decaying into $^{11}B$), the leakage is
taken into account through the reaction $^{11}C(n,2\alpha)^{4}He$.

In figures 2 and 3, we compare the results obtained with i) the NACRE
recommended reaction rates (solid lines) and ii) the CF88 ones (dashed lines).
There is no significant difference except for $^{7}Li$ at high $\eta$.
In this range, ${^7}Li$ comes from $D(p,\gamma){^3}He(\alpha,\gamma){^7}Be$
followed by electron capture. 
So changes in the first rate result directly in changes of the final ${^7}Li$ 
yield.
$^{10}B$ presents the largest difference due to the $^{10}B(p,\alpha)^{7}Be$ 
destruction reaction rate. The NACRE rate is several orders of magnitude
higher due to the inclusion (Rauscher and Raimann 1997) of a 10~keV, 
5/2$^+$ resonance.
However, there is no astrophysical and cosmological consequences since $^{10}B$
is essentially of spallative origin (Vangioni-Flam et al 2000).

Figures 4 and 5 present the updated theoretical primordial abundances from $D$ to
$^{11}B$ using the NACRE compilation (mean values and extreme ones).
D and $^{3,4}He$ are almost not affected.
Due to the uncertainty of the $D(p,\gamma)^{3}He$, the $^{7}Li$ abundance
at $\eta > 3$ is affected by a significant error (about 30\% to be
compared to the 42\% one mentionned by Olive et al 2000 deduced from
the Smith et al 1993 analysis).
At $\eta < 3$, the $^{7}Li$ uncertainty is reduced due to improvements in the
derivation of the S factor of the $T(\alpha,\gamma)^{7}Li$ reaction
(Angulo et al 1999). This is the result of the high precision data provided by
Brune et al (1994) and spanning the entire energy range of interest to BBN
nucleosynthesis. Considering only the precise Brune et al. (1994) data would
even reduce the rate uncertainty.
 However, in this
 specific case, with our method, error compensation occurs between the 
 $^{7}Li(p,\alpha) ^{4}He$ and $T(\alpha,\gamma) ^{7}Li$ reaction rates 
as shown
 in table 1. From the same table, one sees that the maximized uncertainties
 are  $\pm$25\%. Consequently on figure 4, the uncertainty is somewhat
  underestimated which does not affect significantly the general conclusions.
 On the contrary, for $\eta > 3$ region, the errors on the reaction rates
 $D(p,\gamma)^{3}He$,
 $^{3}He(\alpha,\gamma)^{7}Li$ add up and therefore the uncertainties are
 not underestimated.

$^{6}Li$ has a particularly large error bar due to the poor knowledge of
$D(\alpha,\gamma)^{6}Li$. The maximum value of the primordial $^{6}Li/H$
(at low $\eta$) which is of the order of 5.\power{-13} may not be out
of reach of future measurements in halo stars; it represents a factor 10 below
the 6/7 value measured at present in old stars ([Fe/H] = -2.3).
The results presented here are in fair agreement with previous calculations
(Thomas et al 1993, Schramm 1993, Delbourgo-Salvador and Vangioni-Flam 1993,
Vangioni-Flam et al 1999).
We confirm that primordial abundances of $BeB$ are negligible even in the most 
favorable case, and spallation remains the main mechanism to produce them in
the course of the galactic evolution.

\begin{figure} 
\epsfig{file=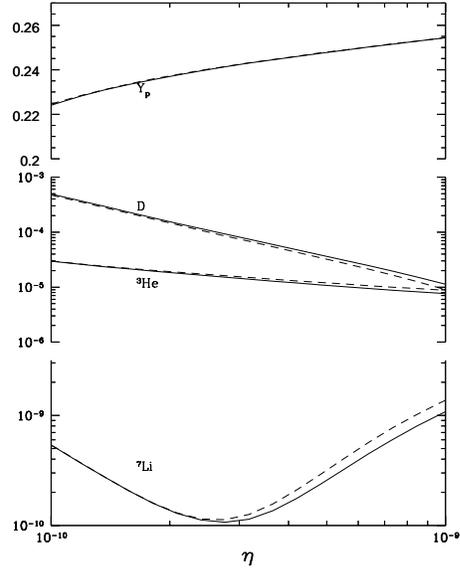,width=8.cm}
\vspace{10pt}
\caption{
Theoretical primordial abundances  of ${^4}He$ (by mass), $D$, ${^3}He$ and 
${^7}Li$ (by number) vs the baryon/photon ratio,  using the reaction rates 
from NACRE (full lines)  and GF88 (dashed lines).}
\end{figure}

\begin{figure} 
\epsfig{file=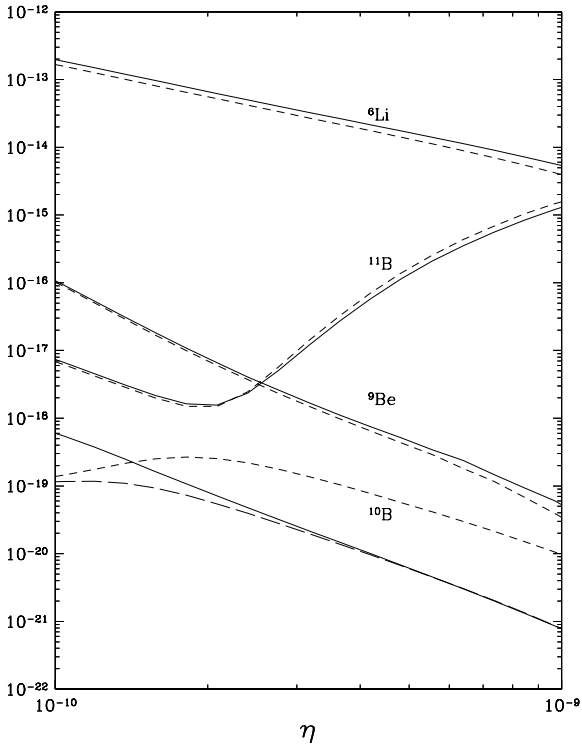,width=8.cm}
\vspace{10pt}
\caption{Theoretical primordial abundances  of ${^6}Li$, ${^9}Be$, ${^{10}}B$ 
and ${^{11}}B$ (by number) vs the baryon/photon ratio, $\eta$, using  the 
reaction rates from NACRE compilation (full lines) and CF88 (dashed lines).
The long-dashed line associated to $^{10}B$ correponds to the Rauscher and
Raimann 1997 evaluation.}
\end{figure}

\section{Astrophysical and cosmological discussion}

\subsection{Astrophysical aspects}
In the following, we decline the astrophysical observational and theoretical
status of each isotope of interest and their possible evolution since the
Big Bang in order to define reasonable error boxes to prepare the 
confrontation to the theoretical calculations.

\subsubsection{Deuterium}

$D$ is particularly sensitive to the baryon/photon ratio, $\eta$, and has
been considered up to now as the best baryometer (e.g. Reeves 1994). However,
due to a certain confusion on $D/H$ abundance evaluations, both at high 
redshifts and in the local Galaxy, some care has to be taken in the 
cosmological use of Deuterium.
Let us present a brief overview of the observations. 

$D$ is measured in three astrophysical and/or cosmological sites i) the local
interstellar medium, ISM ($D_{ISM}$), ii) the protosolar nebula ($D_{ps}$),
iii) the cosmological clouds ($D_{cc}$). 
These three values serve as signposts to follow the evolution of $D$ in the 
Universe and in the Galaxy.
Deuterium, due to its fragility is completely burnt in stars. 
Thus, if no production mechanism is at work, we must have 
$D_{cc}>D_{ps}>D_{ISM}$.

The local $D$ abundance, inferred from UV observations of the nearby ISM,
estimated to $(1.6\pm0.1)$\power{-5} (Linsky et al 1995), is probably not
unique, ranging from 5.\power{-6} to about 2.\power{-5}
(Vidal-Madjar et al 1998, Lemoine et al 1999, Vidal-Madjar 2000).
These variations are lacking explanations.
Thus, there is an ambiguity on the true local $D/H$ value which, by the way, 
serves as a normalisation for the chemical evolutionary models. 
These discrepancies weaken the predictive ability of the evolutionary 
models to derive the primordial $D$ abundance.

$D/H$ ratios are measured in the solar system (Jupiter, Saturn, Uranus, Neptun,
comets). This allows to derive a precise protosolar value of 
$(3\pm0.3)$\power{-5} (Drouart et al 1999), somewhat higher than the estimate
of Geiss and Gloeckler (1998) $(2.1\pm0.5)$\power{-5}.

$D/H$ has also been determined in absorbing clouds on the sightlines of 
quasars.
On one side, Tytler et al (2000) (and reference therein) have found three
absorption systems in which $D/H$ are i) $(3.24 \pm 0.3)$\power{-5},
ii) $4^{+0.8}_{-0.6}$\power{-5}, iii) $<6.7$\power{-5}.
As a fair representation of these data, we adopt the following range :  3.
to 4.\power{-5}. This estimate, if identified with the primordial one,
is unconfortably close to the protosolar one, since it implies a very small
$D$ destruction all along the galactic evolution corresponding to a small
variation of the star formation rate from the birth of the galaxy up to now,
in contradiction with the general trend indicated by the strong increase of
the cosmic star formation rate vs redshift ($0<z<2$) (Blain et al 2000,
Madau 2000).
On the other side, high values of $D/H$ have been reported concerning the 
quasar QSO 1718+4807 ($z$ = 0.701) namely, $D/H = (2.5 \pm 0.5)$\power{-4}
according to Webb et al (1997) and 8.\power{-5}$< D/H <$5.7\power{-4}
according to Tytler et al (2000).
Note that the analysis of Levshakov et al (1999) allowing non gaussian 
velocity distributions leads to lower values.
Consequently, we adopt a second data box bounded by 
8.\power{-5}$<D/H<$ 3.\power{-4}.

\subsubsection{Helium-3}

This isotope is produced in comparable amount to that of deuterium, but at the
opposite, its stellar and galactic story is not simple. Its production and 
destruction are model dependent (Vassiliadis and Woods 1993, Charbonnel 2000).
In spite of great effort directed to its abundance determination in HII 
regions and planetary nebula (Balser et al 1999, Bania and Rood 2000) 
$^{3}He$ cannot be used, at the moment, as a reliable cosmic  baryometer 
(Olive et al 1995, Galli et al 1997) since it is very difficult to extrapolate
its abundance back to its primordial value. 

\subsubsection{Helium-4}

The primordial abundance of $^{4}He$ by mass, $Y_{p}$, is measured in low
metallicity extragalactic HII regions (for a review see Kunth and Ostlin 2000).
In addition to the primordial component, $^{4}He$ is also produced in stars
together with oxygen and nitrogen through global stellar nucleosynthesis.
Therefore, in order to extract the primordial component from the observational
data, it is necessary to extrapolate back the observed $^{4}He$ value down to
zero metallicity.
Olive, Skillman and Steigman (1997) selected 62 blue compact galaxies and
obtained $Y_{p}$ $\sim$ 0.234. Izotov and Thuan (1998) pointed out that the
effect of the HeI stellar absorption has more importance that previously
thought and they reported $Y_{p} = 0.245\pm0.004$.
Recently, Fields and Olive (1998) reanalyzed the observational data and
reported $Y_{p} = 0.238\;\pm\;(0.002)$ stat, $\pm\;(0.005)$ syst where the
errors are 1 $\sigma$ values.
The new determination $Y_{p} = 0.2345\pm0.0030$ by Peimbert and 
Peimbert (2000) on the basis of observations of HII regions in the Small 
Magellanic Cloud points towards the lowest helium value proposed by Fields
and Olive (1998). In this context, two data boxes emerge: 
i) $Y_p = 0.245\pm0.004$ and ii) $Y_p = 0.238\pm0.005$.

\subsubsection{Lithium-6,7}

Recent advances on the determination of $Li$ in halo stars (Spite et al 1996,
Bonifacio and Molaro 1997, Molaro 1999, Smith et al 1998, Ryan et al 1999a)
indicate that the Spite plateau is exceptionally thin ($<$ 0.05 dex).
This small dispersion, together with the presence of $^{6}Li$ in two halo 
stars (Smith et al 1993, 1998, Hobbs and Thorburn 1994, 1997,  Cayrel et 
al 1999) indicate that the stellar destruction of ${^7}Li$ if any, is very 
limited (less than $\sim$ 0.1 dex, see Ryan et al 1999a).
$^{6}Li$ is however of cosmological interest since, being  more fragile 
than $^{7}Li$,  its mere presence in the atmosphere of halo stars 
confirms that ${^7}Li$ is essentially intact in these stars (Vangioni-Flam 
et al 1999, Fields and Olive 1999). This, combined with the very small 
dispersion around the average of the Spite plateau add confidence in 
interpreting it as indicative of the primordial $Li$ abundance (especially 
at the lowest metallicities, where the contamination by spallation is 
expected to be negligible (Ryan et al 1999b).
Stellar modelisation should adapt to this constraint. Indeed, simple models of 
lithium evolution predict little or no depletion (Deliyannis et al 1990),
thus conforting the primordial nature of lithium in metal poor halo stars. 
However, three different mechanisms of alteration of lithium in halo stars
have been suggested i) diffusion/gravitational settling (Michaud 2000 and
references therein), ii) rotational mixing (Chaboyer 1998, Pinsonneault 
et al 1999) and iii) stellar winds (Vauclair and Charbonnel 1995).
Some combinations of these three mechanisms have to be envisioned. 
There is a paradox between the absence of dispersion and the number of the 
processes which could produce a potential dispersion.
This implies either a curious statistical compensation or more radically,
that these physical mechanisms are intrinsically irrelevant. 
However, metallicity independent depletion mechanisms for instance (mixing
induced by gravity waves) cannot be totally excluded (Cayrel, private 
communication).

Consequently, we take the observed value of the Spite plateau including
the observational dispersion ( $A(Li)= 2.2\pm0.04$) to which we add
0.1 dex to account the maximum $Li$ stellar depletion. This corresponds
for the primordial lithium to the following range
1.4\power{-10}$<^7Li/H<$2.2\power{-10}.
This evaluation is in fair agreement with that of Olive et al (2000) but
it is narrower than that of Tytler et al (2000) who have enlarged the limit
of depletion to allow agreement with their low $D/H$ measurement. 
 
In the following, due to the high observational quality and the large sample
analyzed (more than 70 objects), $^{7}Li$ is used as a baryometer, keeping
in mind that this isotope has the pecularity to allow to possible solutions
depending on the side of the lithium valley.

\subsubsection{Beryllium and Boron}

Abundance observations of elemental $Be$ and $B$ in very metal poor stars 
in the halo have made great progresses in the recent years (Gilmore et al 1992,
Duncan et al 1992, Boesgaard and King 1993, Ryan et al 1994, Duncan et 
al 1997, Garcia-Lopez et al 1998, Primas et al 1999, Primas 2000).
These observations concern indeed galactic evolution; the lowest observed 
values (of the order of $10^{-13}$) are much higher than the BBN calculated
abundances.
We confirm that BBN calculated abundances of $^{9}Be$, $^{10}B$ and $^{11}B$
are negligible with respect the measured ones in the more metal poor stars.
The origin of these elements can be explained in term of spallation of fast
carbon and oxygen in the early Galaxy (Vangioni-Flam et al 2000).

\begin{figure} 
\epsfig{file=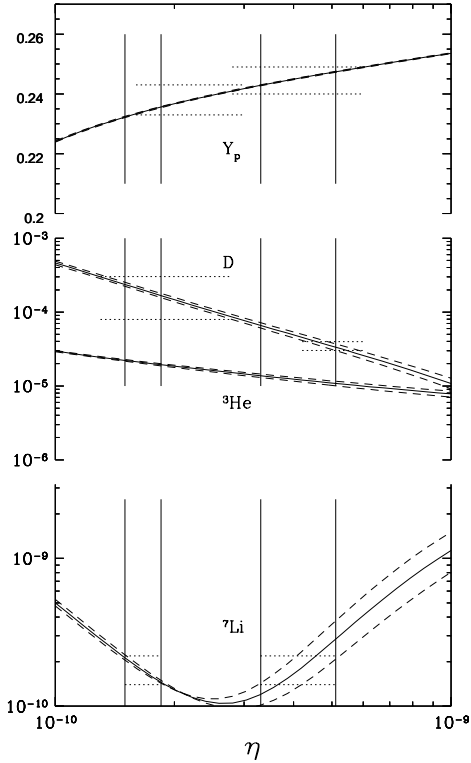,width=8.cm}
\vspace{10pt}
\caption{
Theoretical primordial abundances  of ${^4}He$ (by mass), $D$, ${^3}He$ and
${^7}Li$ (by number) vs the baryon/photon ratio,$\eta$, using the reaction
rates  from the NACRE compilation. 
Full lines: mean values of the reaction rates. Dashed lines: extreme value
of the reaction rates (for details see section 2).
 Horizontal dotted lines indicate the error boxes 
related to different observations, $^{4}He$ left: Fields and Olive 1998, 
$^{4}He$ right: Izotov and Thuan 1998. $D$ left: Webb et al 1997, $D$ right: 
Tytler et al 2000. 
$^{7}Li$: Molaro 1999 and Ryan et al 1999a. Vertical full lines are deduced 
from the error box of $^{7}Li$.}
\end{figure}

\subsection{Baryonic density of the universe}

Once the different error boxes  corresponding to the observed isotopes
abundances corrected for evolutionary effects are established, we can 
compare them to the predictions of the BBN calculations.
As different $D$ and $^{4}He$ measurements are dichotomic, contrary to 
$^{7}Li$, we put emphasis on this last isotope to determine a possible
range of baryonic densities. As shown in figure 4, considering $^{7}Li$ 
alone, two possible ranges emerge: i) $1.5<\eta_{10} < 1.9$, 
ii) $3.3< \eta_{10} < 5.1$. For h = 0.65, we get 
i) $0.013 <\Omega{_B} < 0.019$ and ii)  $0.029 <\Omega{_B} < 0.045$.

The first range is in good concordance with the error boxes related to high
$D$ and low $^{4}He$.
However, the largest measured value of $D/H$ seems excluded
($D/H<$ 3.\power{-4}).
The second one, on the right side of the diagram, is in fair agreement with
a lower $D/H$ (except the lowest measured value, $D/H$ = 3.\power{-5}) and 
a higher $^{4}He$ (except also the highest value 0.25).
At this stage of the analysis, we have to admit that two ranges of baryonic
density have to be into account, only future observations will help to 
remove the ambiguity.

It is worth comparing the baryonic density to that of the luminous matter
($\Omega_{L}$) in the Universe to infer the amount of the baryonic dark matter.
Recent estimates of $\Omega_{L}$ ranges between 0.002 and 0.004 (Salucci and
Persic 1999), which is lower than both $\Omega{_B}$ obtained.
The difference makes necessary  baryonic dark matter.
Focusing on spiral galaxies the amount of luminous matter is estimated
to $\Omega_{LS}$ = $1.44^{+1.55}_{-0.2}$\power{-3} ( Sallucci and Persic
1999). Considering that the dynamical mass of the halo of spiral galaxies is about ten
times higher than that of the disk, the corresponding $\Omega_{HS}$ is about
0.015.
In both cases ( $\Omega_B\approx$0.015 or $\Omega_B\approx$0.04) all the dark
matter in the halo of our Galaxy could in principle be baryonic.
Note that since eight years of searches for microlensing events by the
MACHO and EROS collaborations toward the Magellanic clouds have revealed
only a few events.
The fraction of the halo in the form of dark massive compact objects,
using a typical model, is estimated to only 20\% by Alcock et al (1999) in
agreement with the limits given by Lasserre et al (2000).
On the other hand, the observations of the Lyman alpha forest clouds between
the redshifts 0 and 5, lead to a corresponding $\Omega_B$ of about 0.03 ($\pm 0.01$), taking 
 into account the uncertainty related to ionised hydrogen (Riediger, Petitjean and Mucket 1998).
This value is thought to reflect the bulk of the baryons at large scale.
It is more consistent with our high $\Omega_B$ range.

\begin{figure} 
\epsfig{file=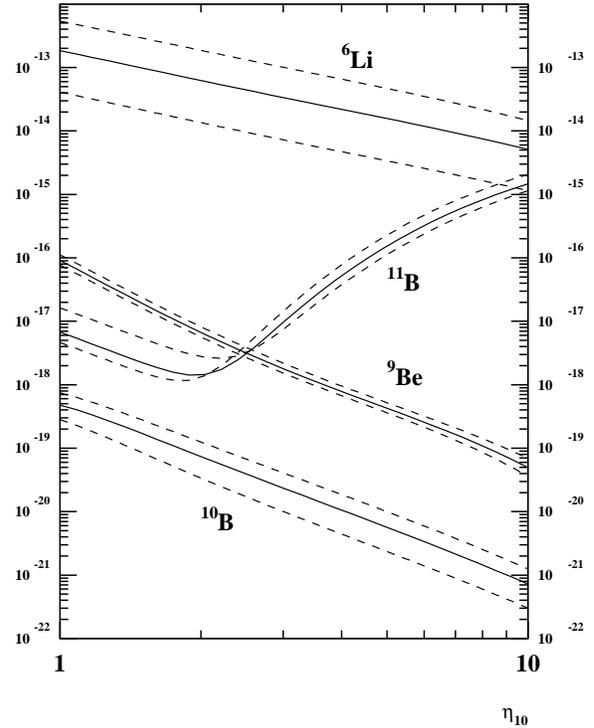,width=8.cm}
\vspace{10pt}
\caption{Theoretical primordial abundances  of ${^6}Li$, ${^9}Be$, ${^{10}}B$ 
and ${^{11}}B$ (by number) vs the baryon/photon ratio, $\eta$, using
the reaction rates from NACRE. Full lines: mean values of the reaction rates.
Dashed lines: extreme limits of the reaction rates.}
\end{figure}

\section{Conclusion}

Big bang nucleosynthesis has been studied since a long time, but it deserves 
permanent care since it gives access to the baryon density which is a key 
cosmological parameter. This work has been aimed at integrating the last 
development in both fields of nuclear physics and observational abundance 
determination of light elements.

1. The update of the reaction rates of the BBN using the NACRE compilation
does not lead to crucial modifications of the general conclusions 
concerning the baryonic content of the Universe. The average values of the 
abundances of isotopes of cosmological interest are in general similar to 
that calculated on the basis of the Caughlan-Fowler (1988) compilation.

2. However, the modification of the $D(p,\gamma)^{3}He$ reaction rate leads to
a $^{7}Li$ abundance slightly lower at $\eta > 3$. But the uncertainty on this 
reaction rate remains high ($\pm$30\%). At $\eta< 3$, the revision
of the $T(\alpha,\gamma)^{7}Li$ reaction rate leads to a very neat reduction 
of the uncertainty of the calculated $^{7}Li$ abundance. However this reaction, together
 with the $^{7}Li(p,\alpha)^{4}He$, remain the main sources of the $^{7}Li$ 
  uncertainty at low $\eta$.

3. The abundance of $^{10}B$ is modified by the new $^{10}B(p,\alpha)^{7}Li$
reaction rate, but there is no cosmological consequences.

4. $^{6}Li$ is affected by the large uncertainty of the 
$D(\alpha,\gamma)^{6}Li$ reaction. 
However, $^{6}Li$ is essentially of spallative origin.

5. Owing to the high observational reliability of the $^{7}Li$ abundance data
with respect to the $D$ data avalaible both rare and debated, we choose it as 
the leading baryometer, since it appears that the primitive $^{7}Li$ is almost
intact in halo stars.
Due to the competition between $T(\alpha,\gamma)^{7}Li$ and
$D(p,\gamma)^{3}He(\alpha,\gamma)^{7}Be$ (e $\nu$) $^{7}Li$, the curve of 
$^{7}Li$ vs $\eta$ presents a valley shape. Consequently, the observational
error box of $^{7}Li$ leads to two ranges of $\eta$ : $1.5<\eta<1.9$ and 
$3.3<\eta<5.1$ (corresponding to $0.013<\Omega_B<0.019$ and 
$0.029<\Omega_B<0.045$ for h=0.65). 
In both cases, all the dark matter in the halo of our Galaxy could be baryonic.
However, only a fraction of 20\% is detected  through microlensing events in 
the direction of the Magellanic clouds.

6. These two $\eta$ ranges are confronted to the other available 
cosmologically relevant isotopes, namely $D$ and $^{4}He$. The first $\eta$
range agrees with a high $D$ and low $^{4}He$ values, the second range is 
in concordance with a low $D$ and high $^{4}He$ values. At present, none of 
these solutions can be excluded.

7. In the future, on the nuclear physics front, it would be important to
(re-)measure the $D(\alpha,\gamma)^{6}Li$ reaction to reduce the
uncertainty on the calculated $^{6}Li$ abundance. High precision
measurements over the full energy range of interest to BBN  of the $D(p,\gamma)^3He$,
$^3He(\alpha,\gamma)^7Be$ and $^7Li(p,\alpha)^4He$ reactions would also
reduce the uncertainty on $^7Li$ abundance calculations.
On the astronomical front, more data on $D$ in absorbing clouds on the 
sightlines of quasars at different redshifts are mandatory to remove the 
ambiguity. However, if the large scale $D$ dichotomy remains, it will be
time to invoke specific mechanisms of $D$ production and destruction like
photodisintegration of $D$ and $^{4}He$ by $\gamma$ ray quasars (blazars)
(Cass\'e and Vangioni-Flam 1997) and/or nuclear spallation. 
The observations of $^{6}Li$ in two halo stars has been a great progress and 
a determination of its abundance in very metal poor stars should be pursued.
It will help to constrain even more stringently the possible lithium depletion
in these stars and confort definitively the primordial status of the Spite 
plateau. Together with nuclear improvements, refined $^{6}Li$ measurements
in very metal poor stars (possibly via the VLT) could perhaps lead us towards
the primordial $^{6}Li$ abundance.
 
We warmly thank Roger Cayrel for illuminating discussions, Jurgen Kiener,
Gilles Bogaert and Carmen Angulo for their comments on nuclear data.

\end{document}